\documentclass{aa}

\usepackage{soul}
\usepackage{color}
\usepackage{txfonts}
\usepackage{placeins}
\usepackage{lscape}
\usepackage{lipsum}
\usepackage{subcaption}
\usepackage{hyperref}
\usepackage{graphicx}
\usepackage[percent]{overpic}

\usepackage{mathtools}
\usepackage{enumerate}
\usepackage{ulem}
\usepackage{amsmath}
\usepackage{rotating}

\def\deg {^\circ}

\newcommand{\xmm}{{\it XMM-Newton}}
\newcommand{\chandra}{{\it Chandra}}
\newcommand{\nicer}{{\it NICER}}
\newcommand{\fermi}{{\it Fermi}}

\def\snr {G118.4$+$37.0}
\def \psr {\mbox{1RXS\, J141256.0$+$792204}}
\def \pdot {\dot P}

\def\msun{{\rm M}_{\odot}}

\graphicspath{{./}{Figures/}}

\begin{document}
\makeatletter
\let\if@linenumbers\iffalse
\let\linenumbers\relax
\let\nolinenumbers\relax
\let\runningpagewiselinenumbers\relax
\let\columnwiselinenumbers\relax
\makeatother

\title{
Multi-wavelength study of the high Galactic latitude supernova remnant candidate \snr\ associated with the Calvera pulsar}

\author{Emanuele Greco\inst{1} \and Michela Rigoselli\inst{2,3} \and Sandro Mereghetti\inst{3} \and Fabrizio Bocchino\inst{1} \and Marco Miceli\inst{4} \and Vincenzo Sapienza\inst{1} \and Salvatore Orlando \inst{1}}

\institute{INAF-Osservatorio Astronomico di Palermo, Piazza del Parlamento 1, 90134, Palermo, Italy \\\email{emanuele.greco@inaf.it} \and INAF-Osservatorio Astronomico di Brera, via Brera 28, I-20121 Milano, Italy \and INAF-Istituto di Astrofisica Spaziale e Fisica Cosmica di Milano, via Corti 12, I-20133 Milano, Italy \and Dipartimento di Fisica e Chimica E. Segr\`e, Universit\`a degli Studi di Palermo, Via Archirafi 36, 90123, Palermo, Italy}

\date{\today}

 \abstract
 {The candidate supernova remnant (SNR) \snr\ (Calvera's SNR), discovered as a faint radio ring at high Galactic latitude and coincident with extended \fermi/LAT $\gamma$-ray emission, is likely associated to the X-ray pulsar \psr\ (Calvera). Previous \xmm\ data hinted at soft diffuse X-ray emission inside the ring but lacked sufficient exposure for detailed characterization.}
 {We aim at determining distance, age and physical conditions of Calvera's SNR and at establishing whether hadronic or leptonic processes dominate its $\gamma$-ray emission.}
 {We obtained new \xmm\ observations, and produced count-rate images,
  equivalent width and median photon energy maps to identify optimal regions for spectral analysis. We complemented these observations with a reanalysis of \fermi/LAT $\gamma$-ray data and new Telescopio Nazionale Galileo observations aimed to search for H$\alpha$ emission.}
 {The X-ray diffuse emission is well described by a model of shock-heated plasma with temperature $kT \sim 0.15$ keV, mildly under-solar N and O abundances, and densities $n_e=$0.1--0.7 cm$^{-3}$. According to our estimates, Calvera's SNR is 10--20 kyr old and lies at a distance of 4--5 kpc. A distinct “Clump” region shows harder emission equally well described by a thermal ($kT \sim 1.7$ keV) or a nonthermal model ($\Gamma \sim 2.7$). The brightest X-ray area is close to the $\gamma$-ray peak and an isolated H$\alpha$ filament.}
 {\snr\ is a middle–aged remnant which expands in a tenuous medium and encountered a denser phase, likely the relic of the wind activity of the massive progenitor star.
 The estimated SNR distance is consistent within the uncertainties with that estimated for Calvera, confirming that this peculiar pulsar was born in the explosion of a massive star high above the Galactic disk. Our measured ambient density, together with the patchy morphology of the $\gamma$-ray emission and the detection of H$\alpha$ filaments indicates that  a hadronic origin is compatible with the $\gamma$-ray flux, though a mixed leptonic–hadronic cannot be excluded.}
\titlerunning{Calvera's SNR}
\authorrunning{Greco et al.}

 \maketitle

\section{Introduction}
\label{sect:intro}

The large majority of Galactic supernova remnants (SNRs) are located in the Galactic plane \citep{gre25}, consistent with their origin in massive stars with a typical scale height of $\sim 40$ pc \citep{ufm14}. However, in recent years a few remnants at high latitudes, $|b|>10\deg$, have been discovered in radio and X-ray surveys \citep[e.g.,][]{bhw21,fdw21}. While this is not surprising for nearby objects, there are a few SNRs that are at distances implying large heights above the Galactic disk \citep{krf17,ckb21,ahq22,abb22}. Possible explanations are that they result from Type Ia supernovae (SNe) or from core collapse of run-away stars.
Independent of their origin, these objects are excellent laboratories to study the physics of SNRs, because they are located in a medium which is expected to be less dense and more uniform than that found in the Galactic disk and thus their evolution should be less affected by the interactions with a inhomogeneous environment.

A particularly interesting case is that of the candidate SNR G118.4+37.0 (hereafter also indicated with Calvera's SNR), first discovered with LOFAR as a faint, remarkably circular ring with a diameter of about one degree and total flux density of 1.1 Jy at 144 MHz \citep{abb22}.
The SNR nature of the radio ring was confirmed by \fermi/LAT data that showed the presence of diffuse gamma rays in the 0.5--500 GeV range \citep{araya23,Xg22}. The origin of such $\gamma$-ray emission is unclear, with \citet{araya23} favoring a leptonic scenario and \citet{Xg22} identifying issues with both the leptonic and hadronic channels. Soft X-ray diffuse emission with a thermal spectrum was detected with \xmm\ from a region inside the radio ring, but it could not be studied in detail due to the low exposure time and off-axis position in the field of view (FoV) \citep{zhi11}.

The X-ray pulsar \psr\ (also known as ``Calvera'', \citealt{rfs08}) lies less than $5'$ from the geometric center of the SNR. A recent refined measurement of its proper motion ($78.5\pm2.9$ mas yr$^{-1}$) away from the ring center, \citealt{rmh24}) has strengthened its association with Calvera's SNR, already suggested by the low probability of a chance coincidence between an SNR and a young neutron star at such a high Galactic latitude ($b=+37\deg$).
Calvera is a 59 ms pulsar detected only in the soft X-ray range. Its spin period derivative $\pdot=3.2\times10^{-15}$ \citep{hbg13} implies, with the usual dipole assumptions, a characteristic age $\tau_c \equiv P/2\dot P=2.85\times10^5$~yr, but the association with \snr\ indicates that the true age is smaller than $\tau_c$, similar to other X-ray pulsars found at the center of SNRs \citep{DeLuca17}. In the absence of emission at other wavelengths, the distance of Calvera can be derived only from the modeling of its thermal X-ray emission. The most recent estimate, based on a magnetized hydrogen atmosphere model able to reproduce well both the spectrum and pulse profile, gives a distance of about 3.3 kpc \citep{mrt21}.
If confirmed, this would place Calvera and its SNR at a height of 2 kpc above the Galactic disc.

In this paper we present a multi-wavelength study of Calvera's SNR based on new X-ray observations obtained with \xmm. We complement the X-ray data with deep optical observations performed at the Telescopio Nazionale Galileo (TNG) and by reanalyzing archival \fermi/LAT $\gamma$-ray data. In Sect. \ref{sect:x-ray_data} we describe the observations and data reduction; image and spectral analysis are illustrated in Sect. \ref{sect:img_analysis}; discussion on the results is presented in Sect. \ref{sect:disc}; in Sect. \ref{sect:conc} we wrap up the results and draw our main conclusions.

\section{Observations and data analysis}
\label{sect:x-ray_data}

\subsection{\xmm}
Our new \xmm\ observation of Calvera's SNR, with a planned exposure time of 35 ks, was centered slightly to the east of the diffuse X-ray emission discovered by \citet{zhi11} in the 2009 observations pointed at the Calvera pulsar. Due to visibility constraints, our data were obtained in three separate pointings carried out between August and October 2024, but the last one (obsID 0940970301) was severely impacted by high background, resulting in no usable science data from the EPIC camera\footnote{A further pointing (obsID 0940970401) was scheduled but never performed because of the extremely high background radiation level}.
The details of the two pointings used in this work are summarized in Table~\ref{tab:xmm_obs}.

\begin{table*}[!ht]
 \centering
 \caption{\xmm\ observations of \snr\ (Calvera's SNR) }
\renewcommand{\arraystretch}{1.2}
 \begin{tabular}{c|c|c|c|c|c|c}
 \hline\hline
 OBS ID & Date & Start & Stop& Camera & Exposure (ks) & Exposure (ks) \\
&& (UT)& (UT)& & Unfiltered& Filtered \\
 \hline
 &&&&MOS1 &8.9 & 8.3\\
0940970101 & 2024-08-31 & 10:11:19 & 17:04:39 &MOS2&8.9 & 8.5 \\
 &&&&pn& 9.6 &5.7\\
 \hline
 &&&&MOS1& 14.0& 3.7\\
0940970201 &2024-09-02 & 11:38:11 &17:16:31 &MOS2& 14.0& 4.5\\
 & &&&pn& 13.0& 1.6\\
\hline
 \end{tabular}
 \label{tab:xmm_obs}
\end{table*}

We performed data reduction and generated images, exposure maps, and spectra using version 21.0.0 of the Science Analysis System (SAS) software. To mitigate contamination from soft proton flares, we filtered out time period of high background using the \texttt{espfilt} task.
While obs.ID 0940970101 was largely unaffected, the effective exposure times for obs.ID 0940970201 were reduced by factors of approximately 3 for the MOS and 10 for the pn (see the last column of Table~\ref{tab:xmm_obs}).

Throughout the analysis, we included only events with \texttt{FLAG} = 0 and \texttt{PATTERN} < 13 for MOS and < 5 for pn data.
We used the SAS task \texttt{emanom} to assess whether any EPIC MOS chips exhibited anomalously high quiescent particle background levels below 1 keV. Since most of the emission in the target region lies below 1 keV, we excluded the following chips from our analysis: CCD 4 in MOS1 and CCD 5 in MOS2 for observation 0940970101, and CCDs 4 and 5 in MOS1 and CCD 5 in MOS2 for observation 0940970201.

To generate background-subtracted and vignetting-corrected images, we adopted a double-subtraction method following the procedure outlined in \citet{mbo17}. First, we subtracted the non-photonic background by scaling the Filter Wheel Closed data using the count-rate ratio measured in the unexposed corners of the CCD chips (i.e., areas outside the field of view). Then, we subtracted the photonic background from the resulting photonic-only data.

We combined the background-subtracted count images from each observation using the \texttt{emosaic} task and applied a correction to account for the differing effective areas of the pn and MOS detectors, thereby producing MOS-equivalent mosaicked images.\footnote{The scaling factor for the pn images was estimated using WEBPIMMS: \url{https://heasarc.gsfc.nasa.gov/cgi-bin/Tools/w3pimms/w3pimms.pl}} Exposure maps were generated for each energy band, and a combined mosaicked exposure map was created to normalize the count images. The resulting background-subtracted, vignetting-corrected, and exposure-corrected count-rate image was then adaptively smoothed using the \texttt{asmooth} task to achieve a signal-to-noise ratio of 25. In all images, north is up and east is to the left.

To verify that the X-ray excess detected at large off-axis angle by \citet{zhi11} was not caused by unresolved point sources, we carried out a source detection using the SAS task \texttt{edetect\_chain}.
We initially used background-subtracted count-rate images in the 1.2 to 2.5 keV band, where diffuse emission is negligible, to identify and exclude hard point-like sources. We then ran the source detection procedure simultaneously across multiple soft energy bands to increase sensitivity.
Several candidate sources were detected, including some within the diffuse X-ray excess region. A thorough cross-identification with known sources was therefore required to assess the true nature of the detected emission, and it is presented in detail in Appendix \ref{app:pointsources}.

For the spectral analysis, we corrected for vignetting using the \textit{SAS} tool \texttt{evigweight}, and extracted spectra with \texttt{evselect}. Response matrix files and ancillary response files were generated using \texttt{rmfgen} and \texttt{arfgen}, respectively. The spectra were grouped to contain a minimum of 25 counts per bin, enabling the use of $\chi^2$ statistics. Spectral analysis was performed with XSPEC (version 12.14.1; \citealt{arn96}), by simultaneously modeling the MOS and pn spectra from both observations. Chemical abundances are defined as A$_{\rm{i}} = \frac{\rm n_i}{\rm n_H}$ \large($\frac{\rm n_H}{\rm n_i}$\large)$_\odot$, where n$_{\rm H}$ and n$_{\rm i}$ are the hydrogen number density and the ion number density of the chemical species i, respectively. We adopted the \texttt{lpgs} abundance table from \citet{lpg09}. To validate the robustness of our results, we extracted background spectra from two regions located near the diffuse X-ray excess. All quoted uncertainties on spectral parameters correspond to the 90\% confidence level.

\subsection{\fermi/LAT}
We analyzed nearly 17 years of \fermi/LAT data (from 2008-08-04 to 2025-04-28) between 1--500 GeV,
using the Pass 8 data processing (P8R3) with the \texttt{fermitools} (v2.2.0) and \texttt{fermipy} (v1.3.1) packages.
We selected the Pass 8 `source' class and `front+back' type events coming from zenith angles $<90\deg$ and from a circular region of interest (ROI) with radius of $7\deg \!.5$ centered at R.A. $= 212\deg \!.8$, Dec. = $79\deg \!.4$.

We binned the data with a spatial scale of $0\deg \!.05$ per pixel and ten bins in energy for exposure calculations.
The instrument response function \texttt{P8R3-SOURCE-V3} was used.
We included all the sources from the 4FGL-DR4 catalog \citep{bbb23} within $10\deg$ from the center of the ROI, the Galactic (\texttt{gll-iem-v07}) and the isotropic (\texttt{P8R3-SOURCE-V3-v1}) diffuse components. We allowed the spectral normalizations of all these components free to vary.

Within the extent of Calvera's SNR a dim, point-like source named 4FGL J1409.8+7921 is found (test statistic TS $= 17.0$), which is well fit by a power-law spectral model with photon index $\Gamma=1.9\pm0.2$ \citep{bbb23}.
As already pointed out by \citet{araya23,Xg22}, this source is better described by an extended morphology. We removed it from the list of the cataloged sources and fit the data. Significant residuals extending for about $1\deg$ are found.
The TS map shown in Fig.~\ref{fig:fermi_TS} has been obtained modeling these residuals with point-like sources having a power-law spectrum ($\Gamma=1.9$, but we obtained similar results also for $\Gamma=1.8$ and $\Gamma=2.0$).

\begin{figure}[!ht]
 \centering
 \includegraphics[width=0.9\linewidth]{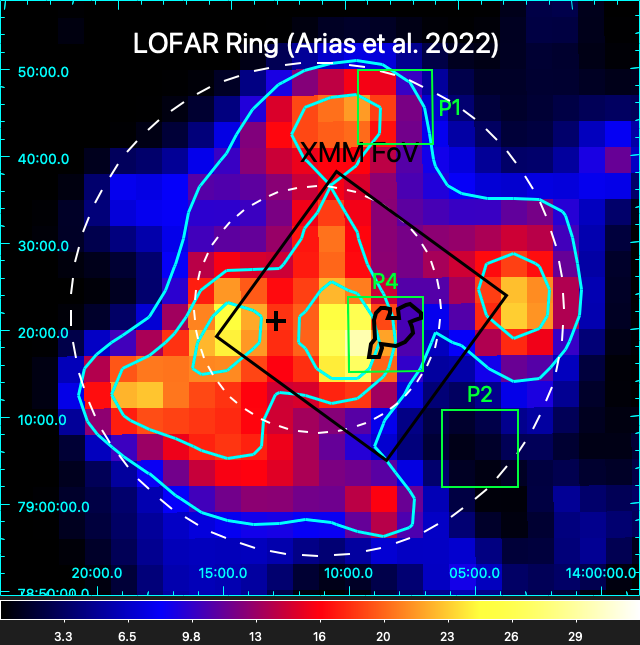}
 \caption{Detection significance (TS) map from \fermi/LAT data in the 1--500 GeV band. The cyan contour levels indicate regions with TS values of 9, 16 and 20 (corresponding to $3\sigma$, $4\sigma$ and $4.5\sigma$, respectively). The white dashed annulus marks the inner and outer boundaries of the radio emission detected by \citet{abb22}, the black box indicates the field of view of our \xmm\ observations while the green boxes mark the field of view of the TNG/DOLORES pointings P1, P2 and P4. The position of the Calvera pulsar is marked with the black cross while the region of diffuse X-ray emission is indicated by the black line. }
 \label{fig:fermi_TS}
\end{figure}

\subsection{TNG/DOLORES}
We designed a deep optical campaign to detect Balmer-dominated emission from the forward shock in Calvera's SNR and possibly radiative faint optical emission associated to shock expansion in a dense medium. We used the DOLORES (Device Optimized for the LOw RESolution) imaging and spectroscopic camera at the 3.6m Telescopio Nazionale Galileo (TNG) at Roque del Los Muchachos in Canary island of La Palma, Spain. The log of the observations is reported in Table~\ref{tab:lrs_obs}.

\begin{table*}[!ht]
 \centering
 \caption{TNG/DOLORES optical observation log}
\renewcommand{\arraystretch}{1.2}
 \begin{tabular}{c|c|c|c|c}
 \hline\hline
 Observation &Pointing coordinates & Mode/Filter & Date & Exposures \\
 \hline
P1 & $14^{\rm h} 07^{\rm m} 48.00^{\rm s}$, $+79\deg 46' 40.0''$ & Imaging H$\alpha$ \#22 \& R \#12 & 2024-05-10 & $12\times 30$m \\
P2 & $14^{\rm h} 04^{\rm m} 36.00^{\rm s}$, $+79\deg 07' 03.0''$ & Imaging H$\alpha$ \#22 \& R \#12 & 2024-05-11 & $14\times 30$m \\
P4 & $14^{\rm h} 08^{\rm m} 22.22^{\rm s}$, $+79\deg 20' 18.7''$ & Imaging H$\alpha$ \#22 \& R \#12 & 2025-04-28 & $14\times 30$m \\
\hline
 \end{tabular}
 \label{tab:lrs_obs}

\end{table*}

DOLORES in imaging mode has a FoV of $8.6^\prime\times 8.6^\prime$. The images have been reduced and stacked using standard procedures implemented in the IRAF software package and AstroLib IDL library. Flux calibration was obtained by observations of the spectrophotometric standard GRW+70 5824, obtained in the same nights of the science observations and reduced in the same way. The $5\sigma$ sensitivity reached in our observations is $2.5\times 10^{-17}$ erg cm$^{-2}$ s$^{-1}$ arcsec$^{-2}$.
We obtained a continuum-free H$\alpha$ image for each pointing by subtracting the image taken with the narrow band H$\alpha$ \#22 filter (with a central wavelength of 657.5 nm and a FWHM of 6.1 nm) and a rescaled version of the broad band R \#12 filter (central wavelength of 644 nm and FWHM of 148 nm).

\section{Results}
\label{sect:img_analysis}

\subsection{X-ray image analysis}

Fig.~\ref{fig:count-rates} shows the \xmm\ count-rate images of Calvera's SNR obtained as described above in a soft band (0.3--1 keV, left) and a hard band (1--2.5 keV, right). The most prominent feature is a bright diffuse structure in the soft band near the center of the field of view (outlined by the white polygon), that can be divided into three regions, as indicated by the green lines. The region labeled Central is the brightest one, while those named South and North are fainter, yet presenting significant emission with respect to the surroundings. This diffuse feature nearly vanishes in the hard band, indicating the extremely soft nature of the emission.

\begin{figure*}[!htb]
 \centering
 \includegraphics[width=0.45\linewidth]{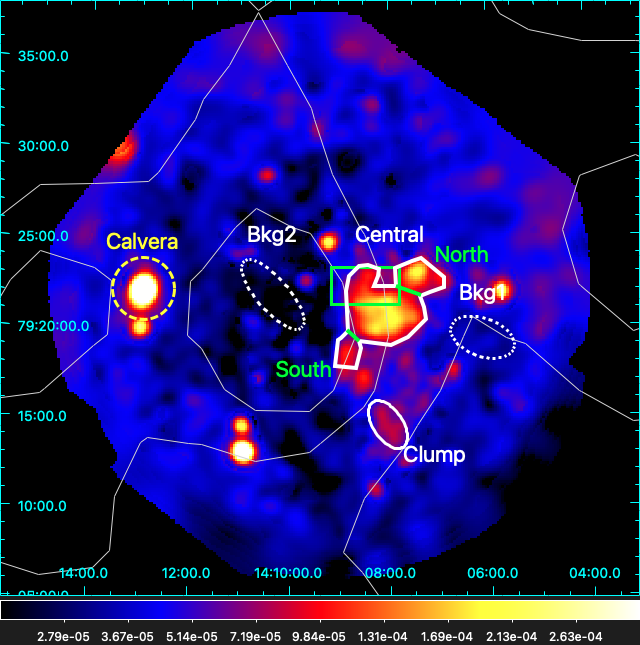}
 \includegraphics[width=0.45\linewidth]{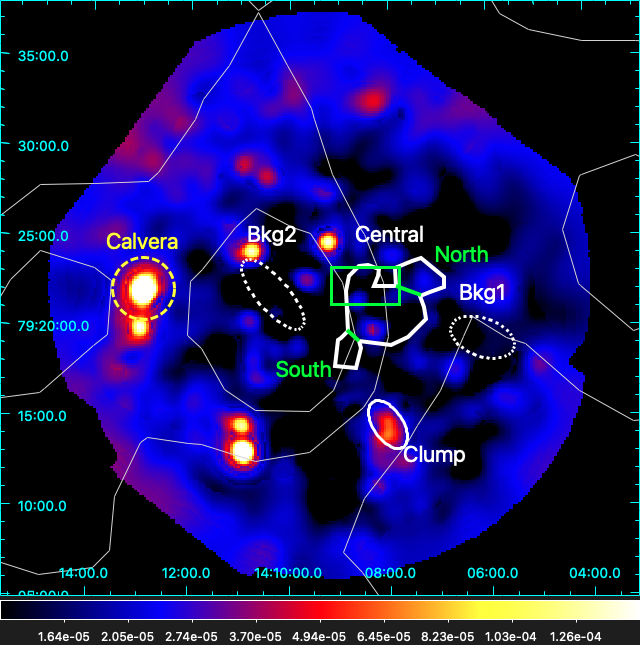}
 \caption{Background-subtracted, vignetting corrected and mosaicked count-rate images of Calvera's SNR in the 0.3--1 keV (left) and 1--2.5 keV (right) energy bands. The area used for the spectral analysis of the brightest diffuse emission (divided into North, Central and South regions) is identified with the white polygon. The white ellipse marks the Clump region and the white dashed ellipses are the two regions used for the background. Calvera pulsar is shown with a yellow dashed circle. \fermi/LAT contours from Fig. \ref{fig:fermi_TS} are overlaid in gray. The green box marks the TNG/DOLORES area of pointing P4 in which the smudge is detected (see Fig.~\ref{fig:smudge}).
 }
 \label{fig:count-rates}
\end{figure*}

\citet{zhi11} reported the detection of the O VII emission line in the spectra extracted from the diffuse X-ray excess.
However, their measurement was limited by the low statistics resulting from the off-axis position in the FoV and the lack of pn data. To better investigate this feature with our new data, we constructed an Equivalent Width (EW) map of the O VII line. For each pixel, the EW was computed as

\begin{equation}
\label{eq:EW}
\rm{EW_{O}} = \frac{\rm{F_{O}} - \rm{F_{O,utl}}}{\rm{F_{O,utl}}},
\end{equation}

\noindent
where $\rm{F_{O}}$ is the total flux in the 0.5--0.7 keV energy band and $\rm{F_{O,utl}}$ is the continuum-only (under-the-line) flux in the same band. $\rm{F_{O,utl}}$ was derived by extrapolating a bremsstrahlung spectrum from the 0.3--0.5 keV to the 0.5--0.7 keV range.
The bright areas in the resulting EW map (left panel of Fig.~\ref{fig:EW-MPE}) correspond to regions where the O VII line flux is significantly enhanced with respect to the underlying continuum.
The most striking result is that the whole region of soft diffuse emission is also bright in the EW map, suggesting a prominent oxygen emission line.

The EW map was also instrumental in defining the regions used for spectral extraction. We prioritized regions that were both bright in the EW map and had sufficient photon counts to enable reliable spectral fitting. The Central part of the diffuse X-ray emission offered the best compromise under these constraints. In addition, we selected two regions in the northern and southern sectors, which appear fainter and have lower EW values, but may still be physically connected to the broader structure.

To further distinguish the emission from the shocked plasma from its surroundings, we produced a map of median photon energy (MPE) by computing, for each pixel, the median energy of all detected photons in the 0.3--2 keV energy band \citep{mdb08}. For this, we used only the pn data, given the significantly different instrumental responses of the MOS and pn cameras, and the superior effective area of the pn detector. Each pixel of the resulting MPE map shown in the right panel of Fig.~\ref{fig:EW-MPE}, has a size of 20 arcsec and is smoothed with a gaussian having a $\sigma$ of 2 pixels. The central region appears with a significantly lower MPE compared to the surrounding emission, consistent with its soft spectrum. Notably, the region with the lowest MPE closely matches the area of highest EW, as expected for a diffuse source with enhanced oxygen line emission.
In both panels of Fig.~\ref{fig:EW-MPE} we also overlap the \fermi\ contours derived from Fig.~\ref{fig:fermi_TS}. Remarkably, the peak of the $\gamma$-ray emission is very close to the area with the highest (lowest) EW (MPE).

Another interesting structure, referred to as the ``Clump'' (white ellipse in Fig.~\ref{fig:count-rates}), stands out in both the soft (0.3--1 keV) and hard (1--2.5 keV) bands as well as in the MPE map. This region has an MPE even higher than that of the Calvera pulsar, yet it appears only marginally brighter than the background in the EW map. These properties, combined with its high surface brightness in the hard band (Fig.~\ref{fig:count-rates}, right panel), suggest that the Clump may be associated with hotter and, likely, more tenuous plasma than that of the other regions.

\begin{figure*}[!htb]
 \centering
 \includegraphics[width=0.42\linewidth]{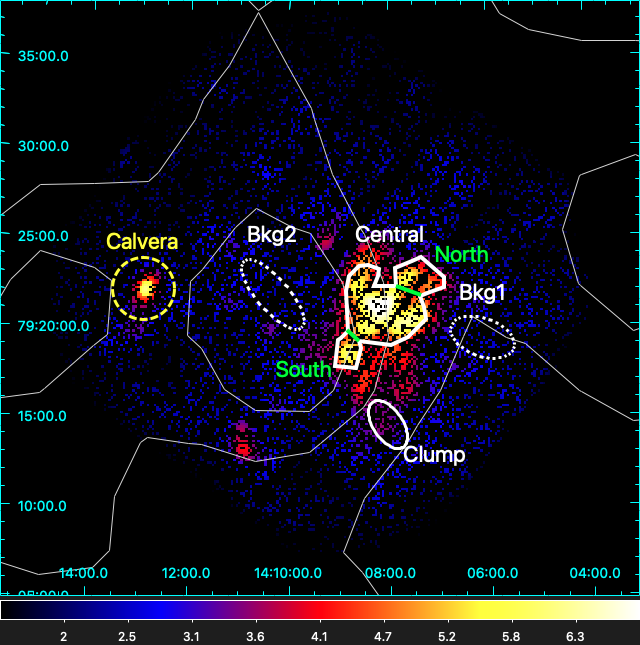}
 \includegraphics[width=0.42\linewidth]{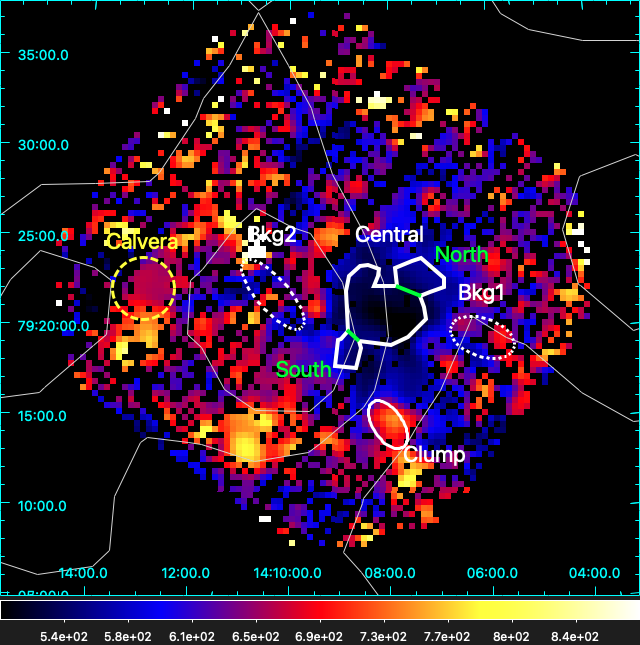}
 \caption{Maps of oxygen EW (left panel) and MPE in the 0.3--2 keV energy band (right panel). The EW map has a pixel-size of 8 arcsec and is adaptively smoothed to a signal-to-noise ratio of 25. The MPE map has a pixel-size of 20 arcsec and is adaptively smoothed with a gaussian having a $\sigma$ of 2 pixels. Values in the color scale are expressed in eV. The regions used for the spectral analysis and the contours of the 1-500 GeV emission are overlaid.}
 \label{fig:EW-MPE}
\end{figure*}

\subsection{X-ray spectral analysis}
\label{sect:spec_analysis}

An important aspect for the spectral analysis is the selection of the background region. Due to the presence of residual soft proton flares and the complex morphology of the field, spatial variations in the background level are present. To mitigate potential biases introduced by relying on a single background region, we adopted two distinct regions — Bkg1 and Bkg2 — chosen to represent the maximum and minimum observed background levels, respectively.

We investigated the nature of the plasma in the selected regions using a simple spectral model composed of a photoelectric absorption component (\texttt{TBabs}) and an emission component from optically thin plasma with variable chemical abundances either in ionization equilibrium (CIE, \texttt{vcie} model) and in non-equilibrium of ionization (NEI, \texttt{vnei} model). Given the limited photon statistics, the column density $N_{\rm{H}}$ was poorly constrained and we could only place an upper limit of $N_{\rm{H}} < 4 \times 10^{20}$ cm$^{-2}$ at the 2$\sigma$ confidence level. Consequently, we fixed $N_{\rm{H}}= 1.7\times10^{20}$ cm$^{-2}$, as determined by \citet{mrt21} from their analysis of \nicer\ spectra of the Calvera pulsar. The best-fit spectral parameters for the different regions are summarized in Table~\ref{tab:xspec_parameters} (uncertainties at the 90\% confidence level). The results are presented using both background regions (Bkg1 and Bkg2) to assess the impact of background selection on the fits. The region ``Total'' represents the sum of the Central, South and North regions. The extracted spectra, along with the corresponding best-fit models and residuals, are shown in Fig.~\ref{fig:spectra} for the Central region and in Fig.~\ref{fig:spectra_clump} for the Clump.

\begin{table*}[!ht]
 \centering
 \caption{Best-fit results for all the regions. }
 \renewcommand{\arraystretch}{1.2}
 \begin{tabular}{c|c|c|c|c|c|c}
 \hline\hline
 Component& Parameter& \multicolumn{4}{c}{Region} \\
 \hline
 & & Total & Central & South & North & Clump \\
 \hline
 & Net counts & $\sim$ 6700 & $\sim$ 5000 & $\sim$ 550 & $\sim$ 1200 & $\sim$ 950\\
 \hline
 \texttt{TBabs} & $N_{\rm{H}}$ (10$^{20}$ cm$^{-2}$) & \multicolumn{4}{c}{1.7 (fixed)} \\
 \hline
 & $kT$ (keV) & $0.165\pm 0.005$& $0.148 \pm 0.007$ & $0.17\pm 0.02$ & $0.191 \pm 0.010$ & $1.7_{-0.4}^{+1.7}$ \\
 & N & / &$0.5_{-0.4}^{+0.5}$ & / &/& / \\
 \texttt{vcie} & O & $0.79_{-0.10}^{+0.12}$ & $0.76_{-0.11}^{+0.15}$ & / & $0.8_{-0.2}^{+0.4}$ & $30_{-20}^{+70}$ \\
 (bkg1) & norm ($10^{-4}$ cm$^{-5}$) & $7.4_{-0.7}^{+0.8}$& $7.0 \pm 1.2$ & $0.45\pm 0.15$& $1.0 \pm 0.3$ & $0.18_{-0.09}^{+0.15}$ \\
 & Flux 0.3--1 keV$^a$& $6.6 \pm 0.3$ & $4.8 \pm 0.3$ & $0.47 \pm 0.07$ & $1.16\pm 0.12 $ & $0.33 \pm 0.05$\\
 & Unabsorbed Flux 0.3--2 keV$^a$ & $8.1\pm 0.4$ & $6.0 \pm 0.4$ & $0.57 \pm 0.08$ & $1.40 \pm 0.15$ & $0.62 \pm 0.09$ \\
 \hline
 & $\chi^2$ (d.o.f.) & 204.80 (181) & 164.5 (149) & 16.68 (18) & 35.79 (39) & 21.86 (27) \\
 \hline
 & $kT$ (keV) & $0.167\pm 0.004$ & $0.151 \pm 0.005$ & $0.17 \pm 0.02$ & $0.192\pm 0.010$ & $1.6_{-0.3}^{+1.0}$ \\
 & N & / &< 0.6 & / & / & / \\
\texttt{vcie}& O & $0.68_{-0.7}^{+0.8}$ & $0.61_{-0.07}^{+0.08}$ & / & $0.8_{-0.2}^{+0.3}$& $25_{-10}^{+27}$ \\
 (bkg2)& norm ($10^{-4}$ cm$^{-5}$) & $8.1\pm 0.7$ & $8.3 \pm 1.1$ & $0.47_{-0.08}^{+0.12}$ & $1.1 \pm 0.3$ & $0.20_{-0.09}^{+0.13}$ \\
 & Flux 0.3--1 keV$^a$& $6.9 \pm 0.3$ & $5.4 \pm 0.2$ & $0.52 \pm 0.06$ & $1.22\pm 0.12 $ & $0.37 \pm 0.05$\\
 & Unabsorbed Flux 0.3--2 keV$^a$ & $8.4\pm 0.3$ & $6.6 \pm 0.3$ & $0.63\pm 0.08$ & $1.48 \pm 0.14$ & $0.68 \pm 0.08$ \\
 \hline

 & $\chi^2$ (d.o.f.) & 201.30 (181) & 174.23 (149) & 18.70 (18) & 47.46 (39) & 28.00 (27) \\
 \hline
 \end{tabular}

 Notes. $^a$Units of 10$^{-13}$ erg s$^{-1}$ cm$^{-2}$
 \label{tab:xspec_parameters}
\end{table*}

\begin{figure}
\centering
 \includegraphics[width=0.9\linewidth]{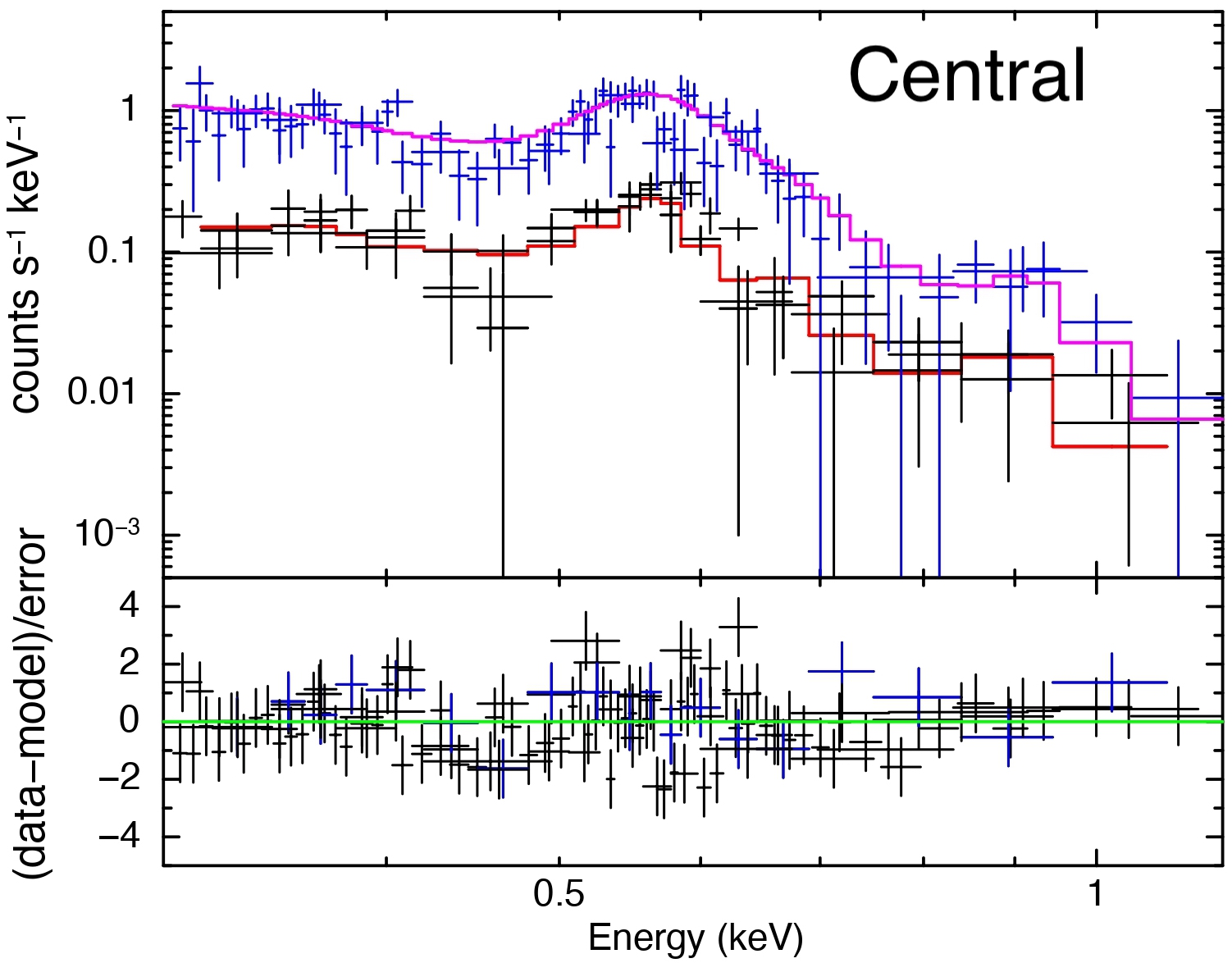}
\caption{EPIC/MOS (black points) and EPIC/pn (blue points) spectra for the Central region fitted with an absorbed CIE component and the corresponding best-fit model (magenta for pn and red for MOS) and residuals.}
\label{fig:spectra}
\end{figure}

\begin{figure}[!htb]
 \centering
 \includegraphics[width=0.45\linewidth]{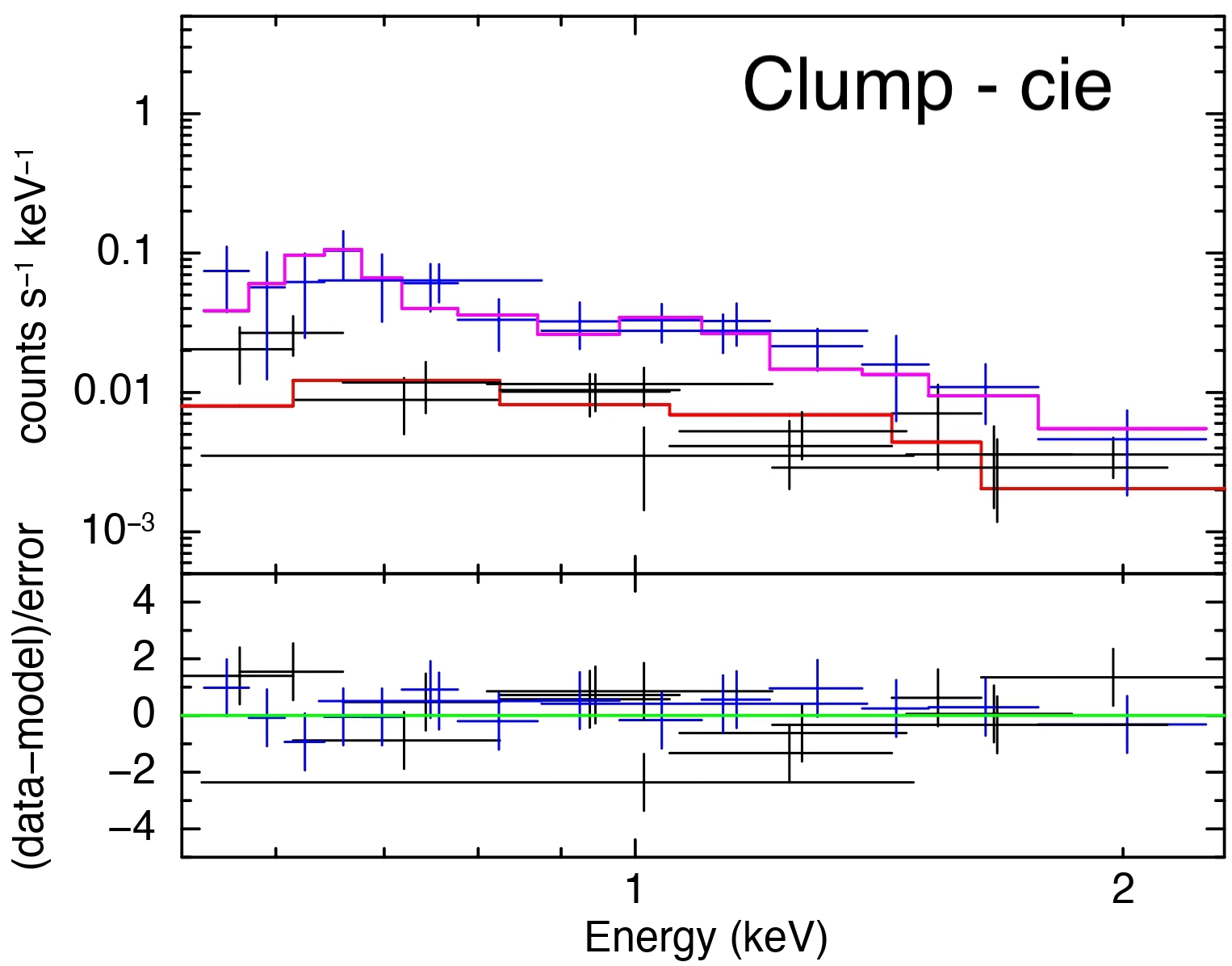}
\includegraphics[width=0.45\linewidth]{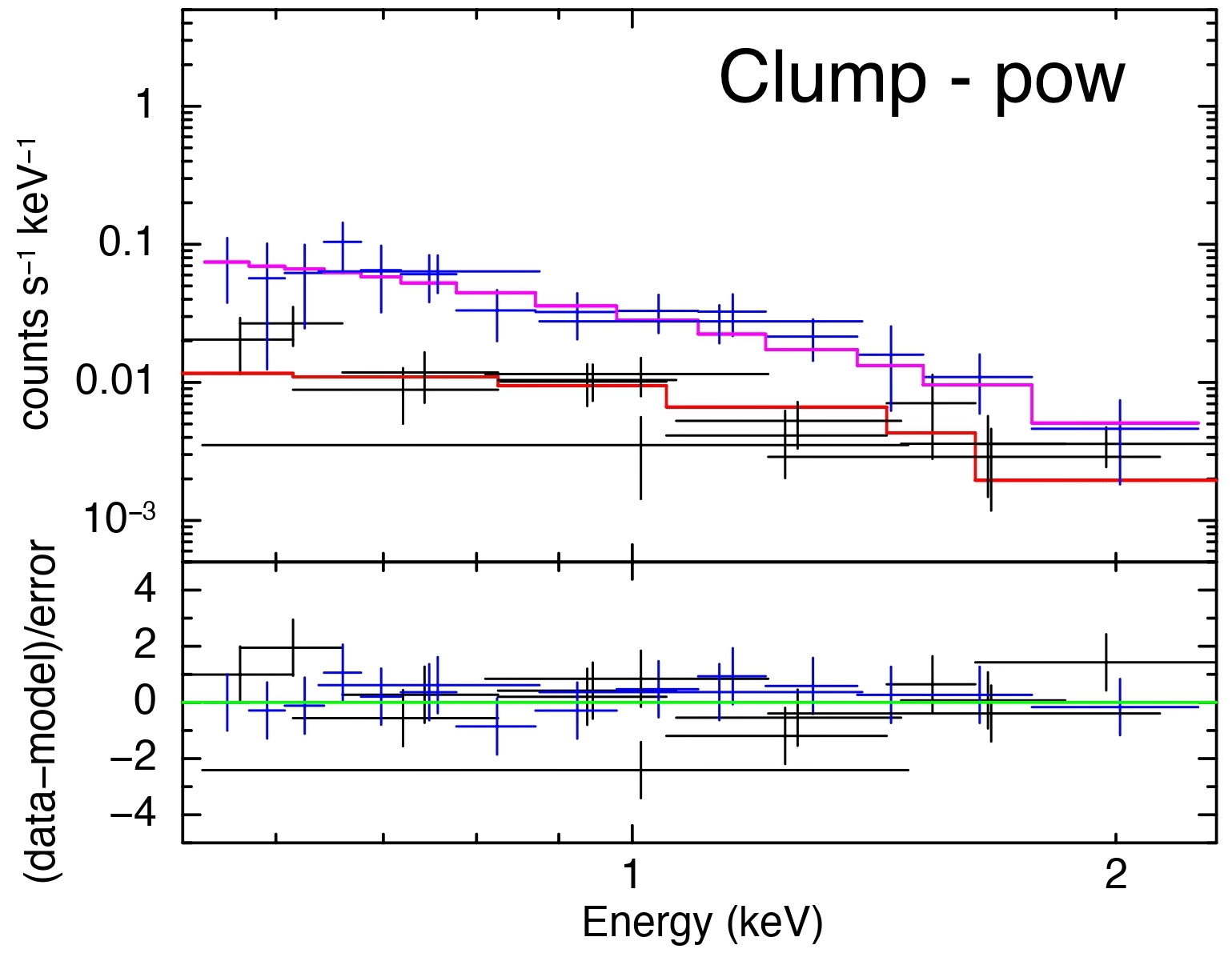}
 \caption{ \textit{Left panel.} Same as Fig.~\ref{fig:spectra} panel but for the Clump. \textit{Right panel.} Same as Left panel but the \texttt{cie} component is replaced by a \texttt{pow} component.}
 \label{fig:spectra_clump}
\end{figure}

The best-fit for the spectrum of the ``Total'' region was found with a temperature of $\sim$ 0.17 keV and by letting the oxygen abundance free to vary.
The spectral parameters derived for the Central, North, and South regions are in good agreement with each other and with those obtained from the Total region. The Central, being the brightest part of the diffuse emission, yielded the most robust results. In this case, letting both the O and N abundances free to vary slightly improved the fit. However, for the North and South regions, the limited photon statistics prevented us from placing tight constraints on elemental abundances — particularly N. Consistent results were obtained when using the Bkg2 region for background subtraction, with only the N abundance in the Central region becoming poorly constrained.\footnote{The lower reduced $\chi^2$ values obtained with Bkg1 are primarily due to its higher flux, leading to increased error bars and correspondingly smaller residuals.}

Overall, the results clearly demonstrate that the Central region offers the tightest constraints on the plasma parameters. Including photons from the North and South regions leads to a dilution of the signal-to-noise ratio, reducing the diagnostic power of the spectral fits. Consequently, we adopt the Central region as the reference for comparing our results with observations at other wavelengths (see Sect. \ref{sect:disc}).

To assess parameter degeneracy, we ran the XSPEC task \texttt{steppar}, producing confidence contours for key parameters. Fig.~\ref{fig:steppars} shows that the O abundance is constrained to be above 0.5 at the 3$\sigma$ confidence level, and the electron temperature lies between 0.14 and 0.16 keV at the same significance. However, the N abundance remains poorly constrained and is compatible with zero at the 2$\sigma$ level.

\begin{figure*}[!ht]
 \centering

 \includegraphics[width=0.45\linewidth]{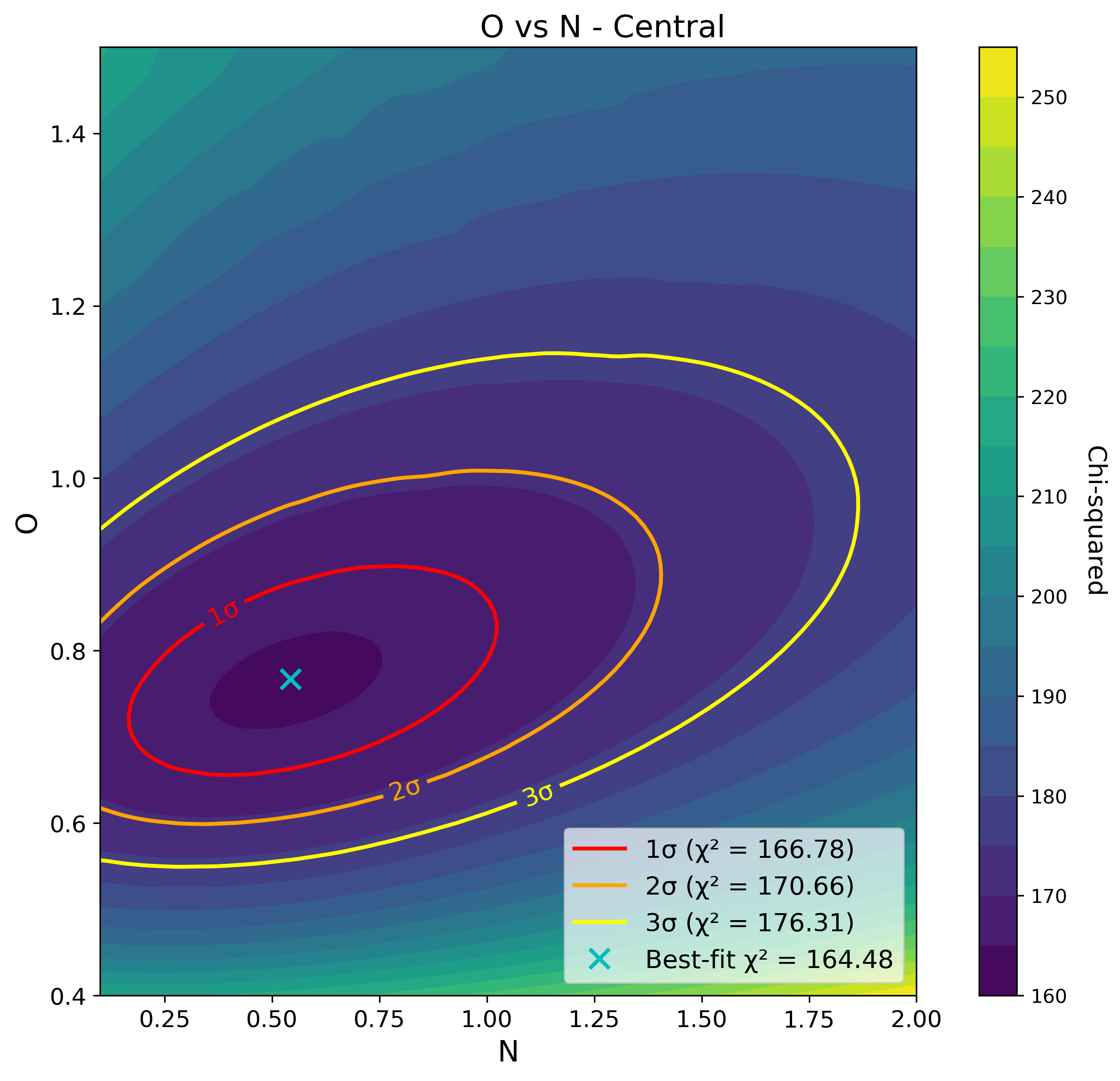}
 \includegraphics[width=0.45\linewidth]{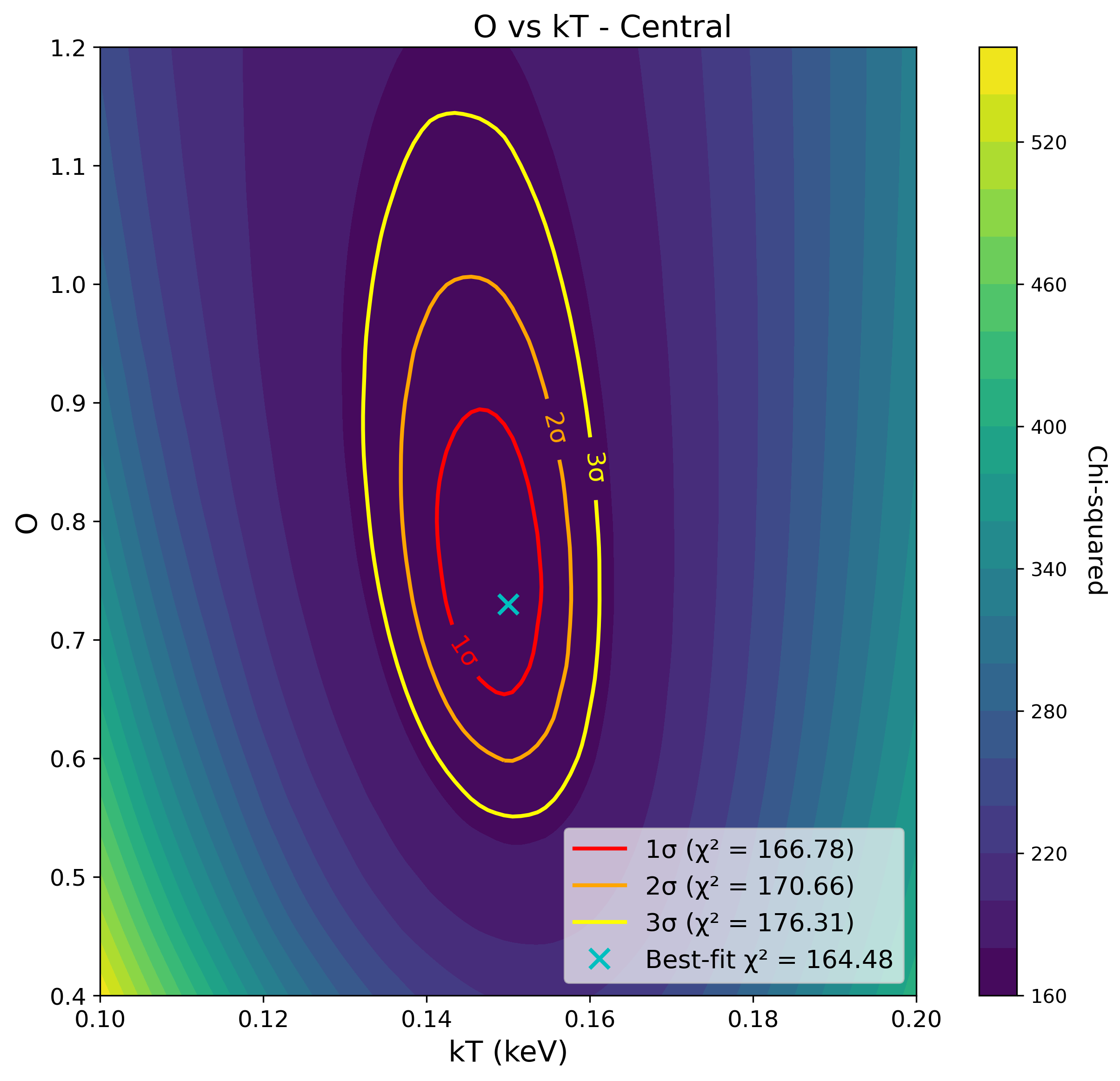}
 \caption{Confidence ranges plots for region Central. Cyan cross marks the minimum (best-fit) in the $\chi^2$ space. Red, orange and yellow contours mark 1$\sigma$, 2$\sigma$ and 3$\sigma$ confidence levels, estimated by considering two parameters of interest. \textit{Left.} Confidence contours for O abundance vs N abundance. \textit{Right.} Confidence contours of O abundance vs electron temperature.}
 \label{fig:steppars}
\end{figure*}

We also tested whether a NEI model (\texttt{vnei}) could improve the fit. Despite introducing an additional free parameter --- the ionization age $\tau$, defined as $\tau = n_e t$, with $t$ the time passed after the encounter with the shock --- we found no significant improvement in fit quality across any of the regions, nor meaningful shifts in the best-fit parameter ranges with the best-fit value of $\tau$ reaching the upper limit of the model.

To better explore the relationship between $\tau$ and $kT$, we generated contour plots (Fig.~\ref{fig:kT-Tausteppar}) analogous to those in Fig.~\ref{fig:steppars}. As commonly observed in X-ray spectra of SNRs, a degeneracy exists between these parameters: acceptable fits can be obtained with lower $\tau$ and higher kT, indicating that NEI conditions cannot be firmly ruled out.

\begin{figure}[!ht]
 \centering
 \includegraphics[width=0.9\linewidth]{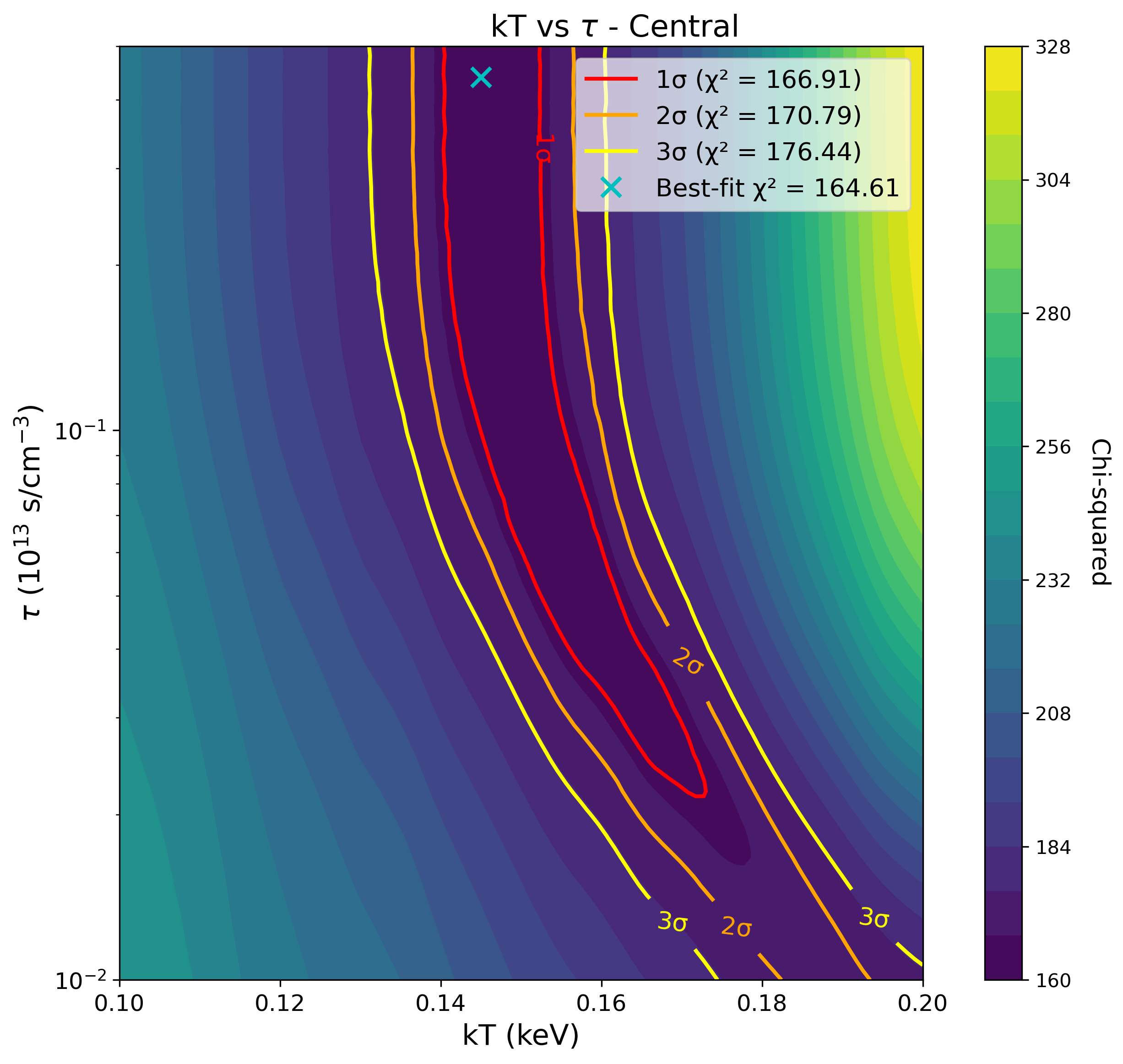}
 \caption{Same as Fig.~\ref{fig:steppars} but for the ionization timescale $\tau$ vs electron temperature $kT$ when adopting a model of plasma out of equilibrium of ionization (\texttt{vnei}). }
 \label{fig:kT-Tausteppar}
\end{figure}

The spectral analysis confirmed that the emission from the Clump region is harder than that of the other regions. In fact, when fitted with the CIE  model, a temperature in the 1.3--3.4 keV range (at the 90\% confidence level) and a high oxygen abundance (>10) were found (see last column of Table~\ref{tab:xspec_parameters}).

We also investigated different models for the Clump, which, in principle, might have an extragalactic origin. Equally good fits to the Clump spectrum could be obtained with a power law of photon index $\Gamma = 2.7 \pm 0.5$ or with a thermal bremsstrahlung with temperature in the range 0.5--1.3 keV (in both cases we fixed the absorption to the total value in this direction, $N_{\rm{H}}$ = $2.4 \times 10^{20}$ cm$^{-2}$, see \citealt{HI4PI}).\\
We also analysed the X-ray spectrum of the Calvera pulsar, finding no significance difference in fluxes or spectral parameters with respect to previous dedicated observations (e.g. \citet{mrt21}).

With the possible exception of the Clump, our
X-ray observations do not reveal any evidence of non-thermal emission.

To set an upper limit on the non-thermal emission above 2 keV, we computed the 3$\sigma$ uncertainty on the 2--10 keV background flux, extracted from the whole \xmm\ FoV after excising all the detected point and diffuse sources.

We then rescaled the obtained value for the ratio between the area of the whole SNR (a circle of 56.8 arcmin diameter)\footnote{We considered a circle instead of a ring because the $\gamma$-ray emission is observed (especially) within the inner radius} and that of the region used for spectral extraction.
For a power law spectrum of photon index $\Gamma = 2.5$, the resulting 3$\sigma$ upper limit is $4.3 \times 10^{-11}$ erg s$^{-1}$ cm$^{-2}$ in the 2--10 keV band.

As a cross-check, we also estimated the same upper limit by adding a power-law component with $\Gamma = 2.5$ to the spectrum of the Central region and fitting it in the 0.3--10 keV range.

We then rescaled the maximum allowed flux ($\Delta\chi^2 = 9$) to account for the differences in the size of the considered regions and in the average count rates (the count rate in the Central region is three times higher than in the rest of the FoV, excluding all point-like sources).
The final upper limit, rescaled on the whole SNR size as defined above, is $2.5 \times 10^{-11}$ erg s$^{-1}$ cm$^{-2}$ in the 2--10 keV band. \\

\subsection{Optical emission}
\label{sect:optical}

We do not detect any diffuse or compact source of H$\alpha$ emission in the shell pointings P1 and P2 of Table~\ref{tab:lrs_obs} (see also Fig.~\ref{fig:fermi_TS}), down to the sensitivity limit of $2.5\times 10^{-17}$ erg cm$^{-2}$ s$^{-1}$ arcsec$^{-2}$. In pointing P4, centered on the Central region (Fig.~\ref{fig:fermi_TS}), the ``smudge'' discovered by \citet{abb22} is clearly detected, as shown in Fig.~\ref{fig:smudge}. We confirmed the line emission nature of this feature and measured an extension of $\sim 1^\prime$, with a complex morphology characterized by two bright ``heads'' at east and a weaker diffuse emission at west. The heads have a surface brightness of 23 mag arcsec$^{-2}$. Further spectroscopic observations are planned in order to assess the radiative or non-radiative nature of this emission, which at present remains an open issue.
Additional faint, diffuse H$\alpha$ features are also visible around the ``smudge'' and indicated in the upper right panel of Fig.~\ref{fig:smudge}; however, these features are too weak to support any firm conclusions beyond noting that they are not associated with any continuum-emitting sources.

\begin{figure*}[!ht]
 \centering
 \includegraphics[width=0.45\linewidth]
 {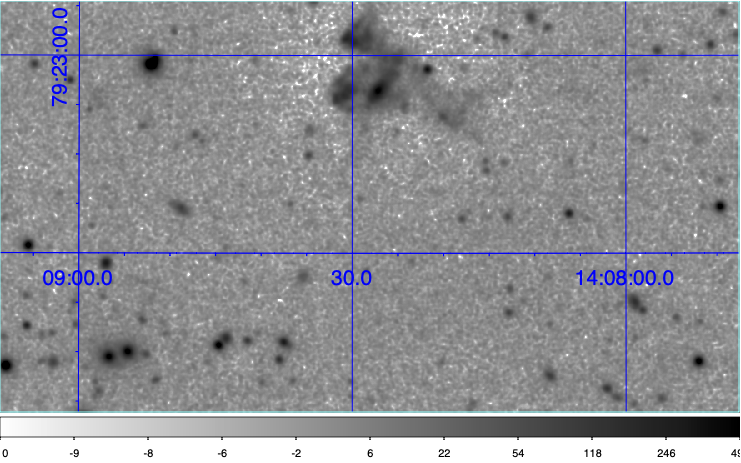}
 \includegraphics[width=0.45\linewidth]
 {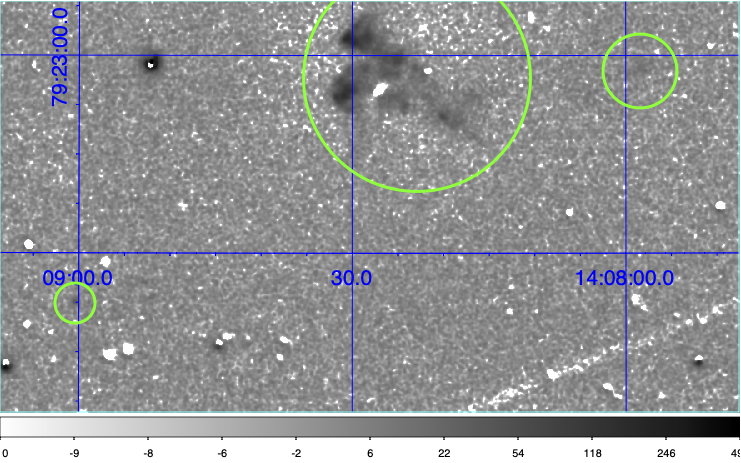}
 \includegraphics[width=0.45\linewidth]
 {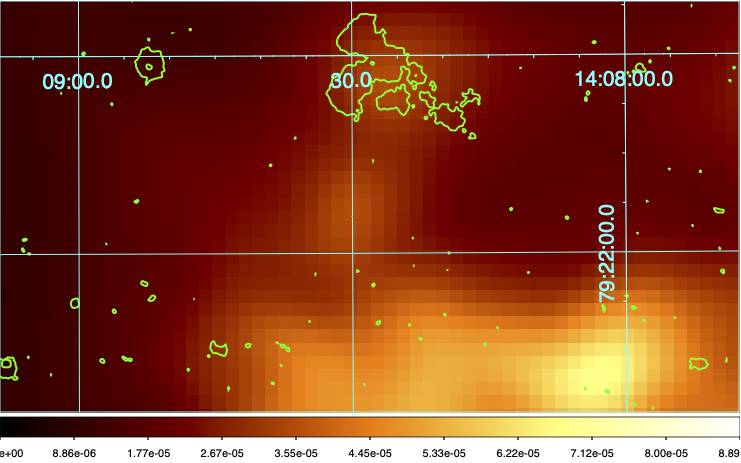}
 \caption{{\it Upper left panel}. TNG/DOLORES deep H$\alpha$ image of the smudge region originally reported by \citet{abb22}, corresponding to pointing P4 (see Table~\ref{tab:lrs_obs} and the green box in Fig.~\ref{fig:count-rates}). The image is background-subtracted and with a logarithmic color scale. {\it Upper right panel}. Same field as the upper left panel, but with additional continuum subtraction using a re-normalized TNG/DOLORES image in the R-band filter. The large green circle has a radius of $34.5$ arcsec and encloses the large and brightest filaments, while the two small circles mark additional faint H$\alpha$ line emission. {\it Lower panel}: X-ray image (same field of view as above) from the left panel of Fig.~\ref{fig:count-rates}, overlaid with H$\alpha$ contours at the level of $1.4\times 10^{-17}$ erg cm$^{-2}$ s$^{-1}$ arcsec$^{-2}$, showing that the bright optical filament has a weak X-ray counterpart within the Central region.}
 \label{fig:smudge}
\end{figure*}

\section{Discussion}
\label{sect:disc}

We presented the analysis of our new \xmm\ observations targeting the diffuse X-ray emission of the candidate SNR \snr, associated to the Calvera pulsar, complemented by \fermi/LAT gamma-ray data and deep H$\alpha$ imaging from the TNG/DOLORES instrument. Through combined imaging and spectral analysis, we found that the diffuse X-rays are well described by thermal emission from a shock-heated plasma with an electron temperature of approximately 0.15 keV. The spectra are characterized by prominent oxygen emission and marginally significant nitrogen features. The plasma appears to be in equilibrium of ionization, although we cannot fully exclude a NEI scenario. No diffuse X-ray emission above $\sim 1.5$ keV is detected in the current data. The proximity between the X-ray emission and the peak of the \fermi/LAT significance map suggests a physical connection between the X-ray-emitting plasma and the gamma-ray source. Moreover, the edge of the diffuse X-ray emission closely aligns with faint H$\alpha$ filaments, including the ``smudge'' structure, while no H$\alpha$ emission was detected in the outer radio shell.

Given the spectral properties of the plasma in the ''Central'' region — most notably the presence of shocked oxygen and, possibly, nitrogen — it is natural to interpret the diffuse X-ray emission as originating from material shaped and enriched by the stellar wind of Calvera's massive progenitor. Therefore, in the following, we refer to this structure as the ``CSM'' (Circumstellar Medium), to emphasize its likely origin as the relic of the progenitor's wind-blown environment, subsequently heated by the passage of the shock. In the following discussion, all the results presented and the analysis performed are focused exclusively on this CSM region, being the one with the most robust constrains.

We can estimate the electron density in the CSM region using the best-fit normalization\footnote{The best-fit normalizations for the fit performed with the NEI and the CIE models are the same} ($7.0 \times 10^{-4}$ cm$^{-5}$) from the spectral analysis (case with bkg1 in Table~\ref{tab:xspec_parameters}) since $\mathrm{norm}=10^{-14}\times \frac{n_e^2 V} {4\pi D^2}$, where $D$ is the distance and $V$ is the volume of the region considered. To estimate the volume we approximated the emitting regions as rotational ellipsoids having the two semi-axis 2.26 and 2.01 arcmin long, with the line-of-sight axis taken as the geometric mean of the two axes projected in the plane of the sky.

To remain as general as possible, we calculated the electron density $n_e$ in the CSM region and the shocked mass for a range of assumed distances between 0.5 and 20 kpc. The resulting electron densities and masses for the CSM region are plotted in the left panel of Fig.~\ref{fig:mass_ne}.

\begin{figure*}[!ht]
 \centering
 \includegraphics[width=0.44\linewidth]{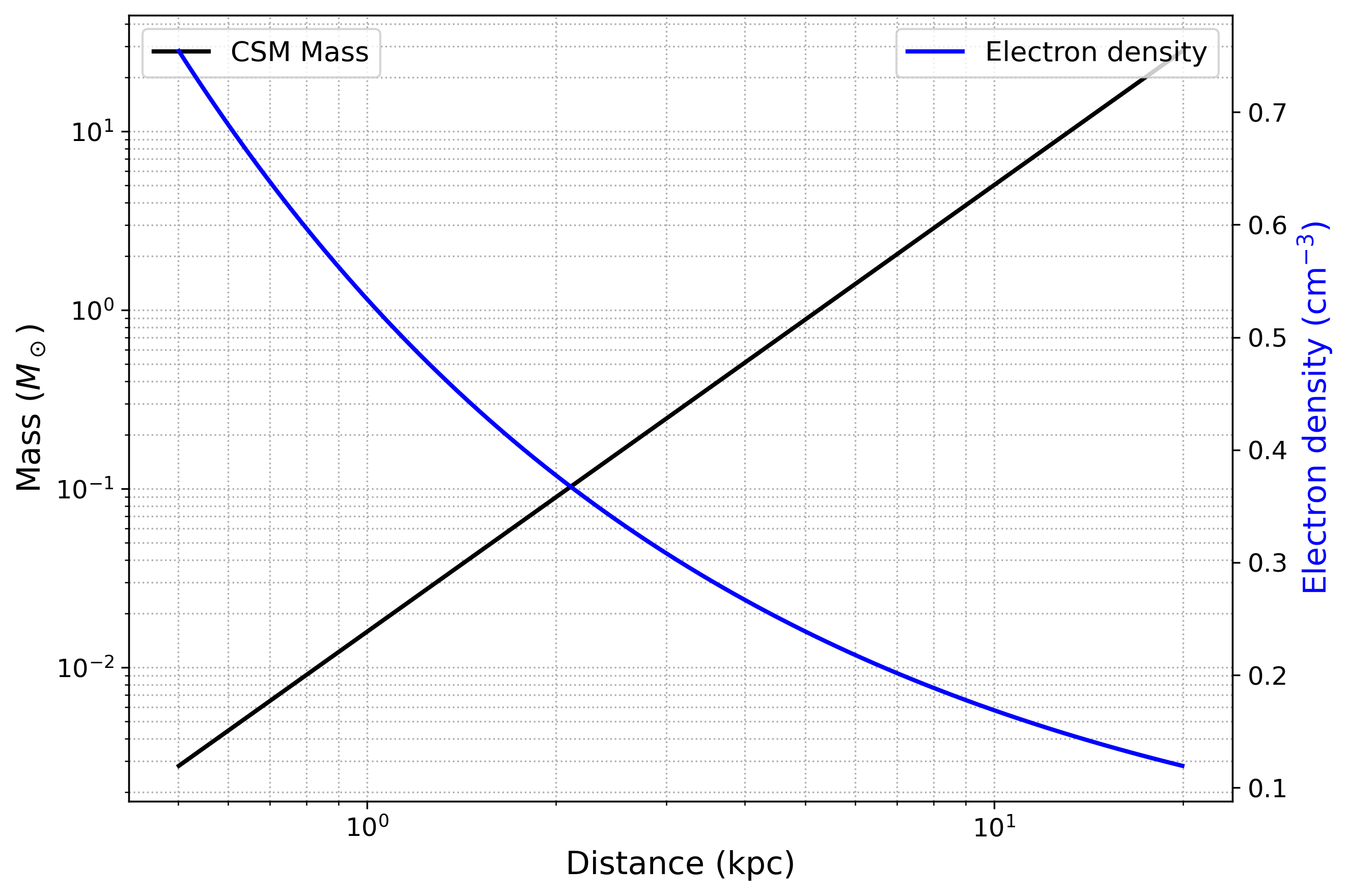}
 \includegraphics[width=0.44\linewidth]{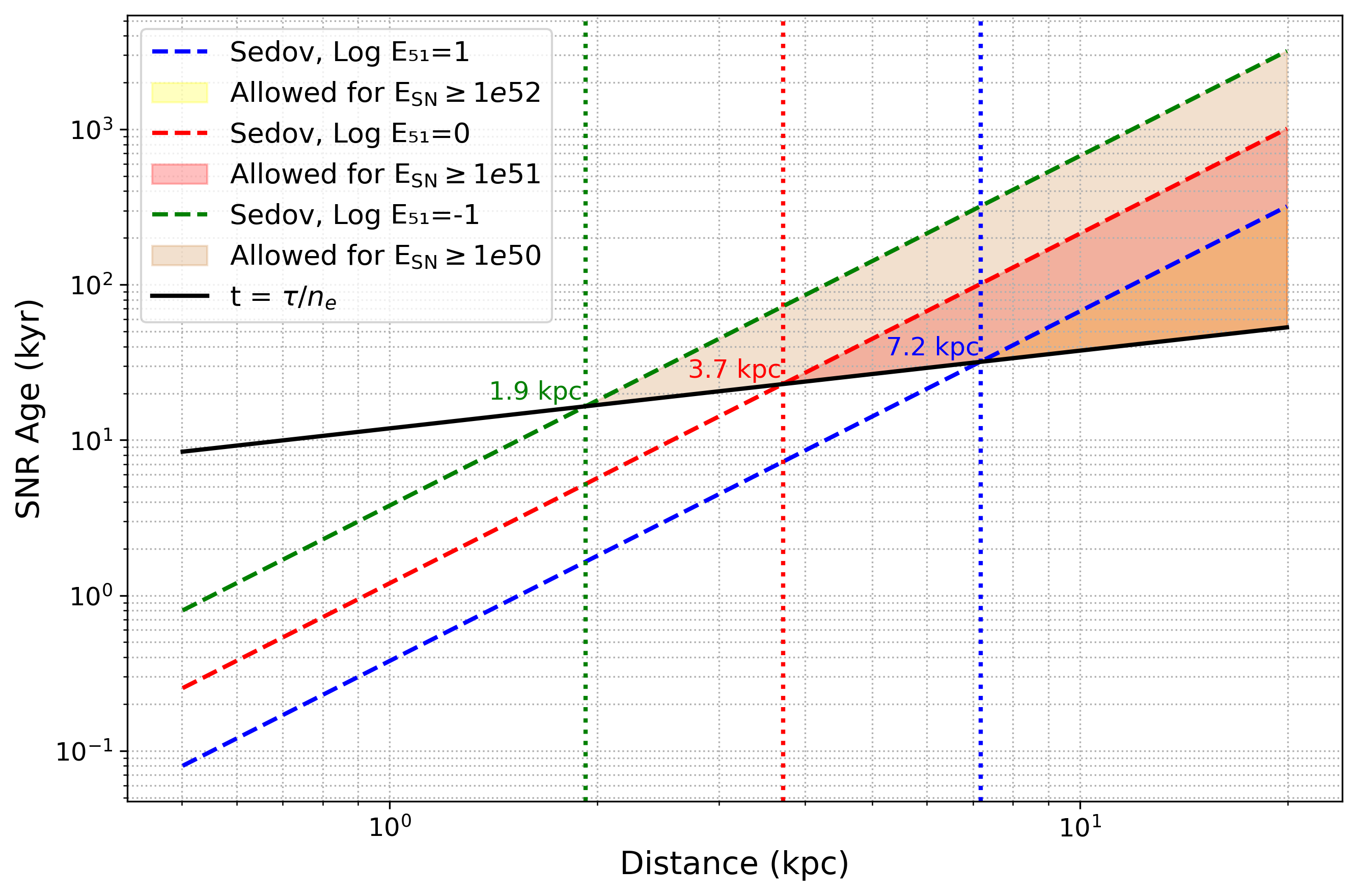}
 \caption{\textit{Left panel.} Inferred shocked mass (black) and electron density (blue) of the CSM region as a function of the distance. \textit{Right panel}. Age of the CSM region estimated adopting the Sedov-Taylor (\citealt{sed59},\citealt{tay50}) solution and assuming density in the LOFAR ring is the same as the one we measured from the X-rays (green, orange and blue dashed lines indicating explosion energies of 10$^{50}$, 10$^{51}$ and 10$^{52}$ erg). Solid black line indicates the age estimated through the $\tau$. Red, yellow and brown area mark the allowed distances, see text for details.}
 \label{fig:mass_ne}
\end{figure*}

The electron density derived in this way falls in the range 0.1--0.7 cm$^{-3}$, significantly higher than the value adopted by \citet{abb22} — $4 \times 10^{-4}$ cm$^{-3}$ — based on the model by \citet{mb13} for the Galactic halo. According to the Sedov-Taylor self-similar solution for SNR evolution \citep{sed59,tay50}, such a low density would imply an age of approximately 7.7 kyr, considering an explosion energy of 10$^{51}$ erg and a distance of 3.3 kpc (\citealt{mrt21}).
This age, however, would correspond to a higher plasma temperature and an earlier evolutionary stage than what is observed in the X-rays, where only the CSM and its immediate surroundings are still emitting. A higher ambient density would better reconcile the age derived from the Sedov-Taylor model with the observational constraints. To test this, we estimated the SNR age $t_{\rm{sedov}}$ using the Sedov-Taylor equations as a function of distance, assuming that the LOFAR ring is characterized by an electron density consistent with that inferred from the X-ray data, and taking advantage of the following equations:
\begin{equation}
t_{\rm sedov} = \left( \frac{R}{\xi} \right)^{5/2} \left( \frac{\rho}{E_{\mathrm{SN}}} \right)^{1/2}
\label{eq:tsedov_def}
\end{equation}
\begin{equation}
\rho = \mu_e m_p \frac{n_e}{4} =  \frac{\mu_em_p}{4} \left(norm\frac{4\pi D^2}{10^{-14} \; V}\right)^{1/2} \propto D^{-1/2}
\label{eq:rho}
\end{equation}

\noindent
where R is the outer radius of the LOFAR ring, $\xi = 2.025$ (e.g. \citealt{vin20}) is the dimensionless factor, $\rho$ is the pre-shock mass density, $E_{\rm{SN}}$ is the explosion energy, $m_p$ is the proton density and $\mu_e=1.2$ is the average mass for each free electron, considering a standard ISM. The electron density $n_e$ is inferred from a post-shock plasma and must be corrected for the compression ratio $r = 4$. Combining Eq. \ref{eq:tsedov_def} and Eq.\ \ref{eq:rho} we obtain the dependency of {$t_{\rm sedov}$} on the distance D,
\begin{equation}
\begin{split}
    t_{\rm sedov} = \left(\frac{R}{\xi}\right)^{5/2} \frac{\mu_em_p}{4} \left(norm \frac{4\pi D^2}{10^{-14} \; V \; \rm{E_{\rm{SN}}}}\right)^{1/2} \propto \\ \propto D^{5/2} D^{-1/4} = D^{9/4}
\label{eq:tsedov_fin}
\end{split}
\end{equation}
The results are shown in the right panel of Fig.~\ref{fig:mass_ne}, for three different explosion energies: $10^{50}$, $10^{51}$, and $10^{52}$ erg.

In the right panel of Fig.~\ref{fig:mass_ne} we also show an alternative estimate of the CSM age. Since the ionization parameter $\tau$ depends only on the electron density $n_e$ and the age $t$, we can estimate $t$ simply by dividing $\tau$ by $n_e$. The main source of uncertainty here lies in the reliability of the $\tau$ measurement. Due to limited statistics, even at the 1$\sigma$ level (see red contour in Fig.~\ref{fig:kT-Tausteppar}), $\tau$ is consistent with values ranging from $2 \times 10^{11}$ s cm$^{-3}$ to $5 \times 10^{12}$ s cm$^{-3}$. To be conservative, we adopt the lower limit of this interval, i.e. $\tau = 2 \times 10^{11}$ s cm$^{-3}$, to estimate the SNR age (black line in the right panel of Fig.~\ref{fig:mass_ne}). The ages which simultaneously satisfy the $\tau$ constrain and are compatible with the Sedov model are shown as shaded areas, with different colors highlighting different explosion energies.

Besides the already mentioned limitations relative to $\tau$ it is important to note that also the assumption that the density in the LOFAR ring is the same as that measured in the CSM region is likely incorrect. Indeed, the observed X-ray, optical and gamma-ray emission suggest a density enhancement in the inner part of the ring, implying that the surrounding medium is comparatively more tenuous. However, we can exploit this difference to improve our constrains on the distance of the source.

In the Sedov model, a larger distance implies a higher shock velocity $v_{\rm sh}$, as the intrinsic size of the remnant would be greater, and this can be seen analytically in:

\begin{equation}
    v_{\rm sh} = \frac{2}{5} \left(\xi\frac{E_{\rm SN}}{\rho}\right)^{1/5} {t_{\rm sedov}^{-3/5}} \propto D^{1/10} D^{-27/20} = D^{-5/4}
    \label{eq:vsedov}
\end{equation}

The shock velocity $v_{\rm{sh}}$ is related to the post-shock temperature $T$ through the relation:

\begin{equation}
    kT = \frac{3}{16} \mu \, m_p v_{\rm{sh}}^2 \propto D^{-5/2}
\label{eq:kt_shock}
\end{equation}

\noindent
where $\mu=0.6$ is the mean particle mass in units of the proton mass $m_p$. Eq. \ref{eq:kt_shock} shows that, for a fixed explosion energy $E_{\rm SN}$, we can compute the expected post-shock temperature as a function of the assumed distance, as shown in Fig.~\ref{fig:ktvsdist}. Since we directly measured the plasma temperature from the X-ray spectra, and we know that the density in the LOFAR ring is lower than in the CSM region, we can constrain the distance by considering only those values that yield a post-shock temperature higher than the observed one. In fact, from Eq. \ref{eq:vsedov} and Eq. \ref{eq:kt_shock} we have: $kT \propto v_{\rm{sh}}^2 \propto n^{-2/5}$. The corresponding allowed regions are shown in green, red and yellow in Fig.~\ref{fig:ktvsdist}, for explosion energies of $10^{50}$ erg, $10^{51}$ erg and $10^{52}$ erg, respectively. By simultaneously taking into account the lower limit on the distance provided by the right panel in Fig.~\ref{fig:mass_ne} and the upper limit above mentioned, we conclude that, in the canonical scenario of a $10^{51}$ erg SN explosion, the SNR is located at a distance of approximately 4--5 kpc. By considering SN explosion energies of $10^{50}$ erg and $10^{52}$ erg the SNR could lie at $\sim$ 2 kpc and $\gtrsim$ 8 kpc, respectively. It is worth mentioning that a distance lower than 1 kpc is unlikely because it would require a mass of shocked CSM $> 10 \msun$ (see left panel of Fig. \ref{fig:mass_ne}), unrealistically high considering that we are looking at a confined region of CSM and not at an entire shell. On the other hand, also distances larger than 10 kpc are very unlikely because they would imply an even greater height above the galactic plane. Finally, we also note that the pulsar distance estimate of 3.3 kpc  \citet{mrt21}  corresponds to  reasonable explosion  energy values of $(7-8)\times 10^{50}$ erg.

\begin{figure}
 \centering
 \includegraphics[width=0.9\linewidth]{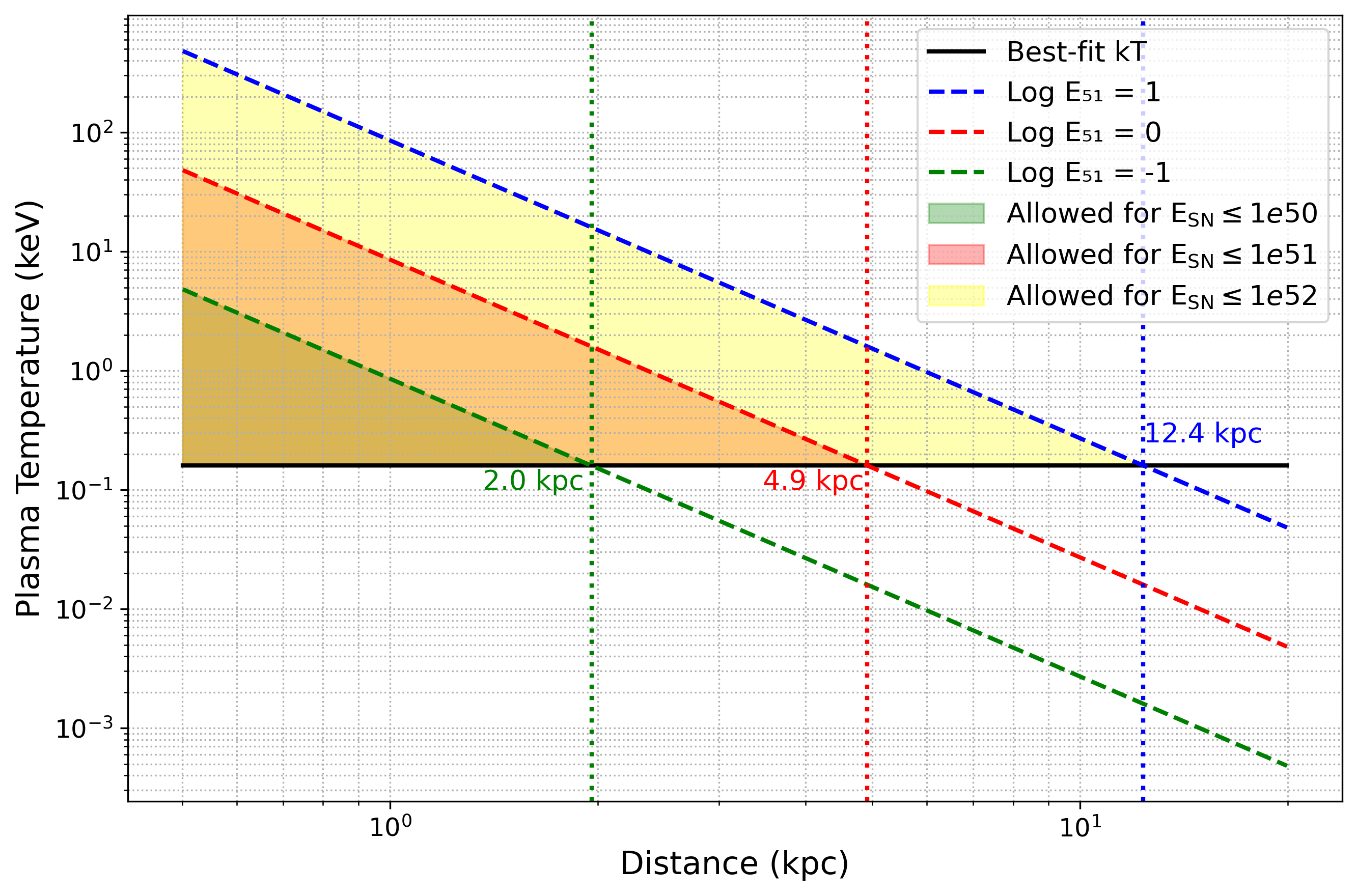}
 \caption{Plasma temperature vs distance inferred by assuming Sedov-Taylor expansion and adopting $kT=\frac{3}{16} \mu \, m_p v_{\rm{sh}}^2$. The black solid line is the best-fit temperature obtained in Sect. \ref{sect:spec_analysis}, the dashed green, red and blue line are the temperature profiles for a SN energy of 10$^{50}$ erg, 10$^{51}$ erg and 10$^{52}$ erg, respectively. The green, red and yellow area mark the allowed distances, see text for details.}
 \label{fig:ktvsdist}
\end{figure}

The observed X-ray emission can be used to place constraints on the multi-wavelength (radio-to-$\gamma$-ray) spectral energy distribution. \citet{araya23} and \citet{Xg22} investigated both leptonic and hadronic scenarios to explain the \fermi/LAT $\gamma$-ray emission from Calvera's SNR. In both studies, a major challenge for the hadronic scenario was the assumed ambient density in which the SNR is expanding: typical Galactic halo values of $\sim 4 \times 10^{-4}$ cm$^{-3}$ (\citealt{mb13}) would require an implausibly high explosion energy of around $10^{54}$ erg. This consideration led both analyses to favor a leptonic origin. However, our X-ray and optical analysis reveals that in the CSM region the density can be significantly higher, in the range of 0.1--0.7 cm$^{-3}$, which is compatible with the values required for a hadronic model. Notably, the location of the X-ray excess is very close to the peak of the \fermi/LAT TS map, as shown in Fig.~\ref{fig:fermi_TS}. This spatial correlation supports the possibility that the TS peak may also be explained by a hadronic origin.

Our upper limit on the non-thermal emission is above the X-ray synchrotron fluxes predicted by the modeling of both \citet{araya23} and \citet{Xg22}. Therefore, we conclude that both leptonic and hadronic processes could plausibly contribute to the observed $\gamma$-ray emission. In particular, it is likely that a combination of the two mechanisms is at work: the peak of the $\gamma$-ray emission coincides with the denser region, where a hadronic origin is more plausible, while inverse Compton scattering may dominate in the outer, less dense regions.

The presence of an optical filaments at the border of the X-ray emission (see Fig.~\ref{fig:smudge}) reinforces the hypothesis of the presence of dense material. A spectroscopic study of the optical emission would be needed to definitely assess the radiative or non-radiative nature of its emission. In any case, it is worth mentioning that the expected morphology of non-radiative H$\alpha$ filaments is quite regular, especially in SNRs propagating in low-density regions (see e.g. \citealt{fnb15} for the high galactic latitude SNR G70.0-21.5). In contrast, the ``smudge'' in Fig.~\ref{fig:smudge} has a more complex patchy morphology which is more typical of bright radiative filaments resulting from slow shocks propagating in local interstellar medium (ISM) enhancement (see e.g. \citealt{bms00} and \citealt{mbr08} for the Vela SNR). The location of the ``smudge'' at the edge of the bright X-ray region (Fig.~\ref{fig:smudge}, lower panel) may indicate that the shock is propagating in a region with increasing density (0.1-0.7 cm$^{-3}$ for the X-ray emitting material and $>1$ cm$^{-3}$ for the optical radiative filament, in analogy to, e.g., what happens in the Vela SNR according to \citealt{bms00}), which is in turn compatible with the possible hadronic nature of the $\gamma$-ray emission.

The association with a still-visible X-ray-emitting SNR indicates that the true age of the Calvera pulsar is significantly lower than its characteristic spin-down age, $\tau_{c}=P/2\dot{P} \sim 285$ kyr. This had also been pointed out by \citet{rmh24}, who inferred an age of $\lesssim10$ kyr based on the proper motion of Calvera.
Such discrepancies between the spin-down and true ages of neutron stars are not uncommon and can be explained in several scenarios: for instance, if the neutron star was born with a spin period close to its current value (the so-called anti-magnetar scenario; \citealt{hg10}), linking Calvera to the class of central compact objects; or if the magnetic field contains strong non-dipolar components that alter the spin-down torque \citep{gh18}.

\begin{table*}[ht]
\centering
\caption{Main characteristics of SNRs at high Galactic latitude. }
\renewcommand{\arraystretch}{1.2}
\begin{tabular}{l|c|c|c|l}
\hline
SNR & Age (kyr) & Distance (kpc) & Pre-shock density (cm$^{-3}$) & Reference \\
\hline
\snr\ (Calvera's SNR) & $\sim 10$ & $\sim 2$ & $0.03-0.2^a$ & This work \\
G249.5$+$24.5 (Hoinga) & $21-150$ & $0.45-1.2$ & $0.16-0.42$ & \citealt{bhw21} \\
G354.0$-$33.5 & $\sim200$ & $< 0.4$ & $\sim 0.1$ & \citealt{fdw21} \\
G116.6$-$26.1 & $\sim40$ & $\sim3$ & $\sim10^{-3}$ & \citealt{ckb21} \\
G70.0$-$21.5 & $\sim 90^b$ & $1-2$ & $< 200$ & \citealt{fnb15} \\
G017.8$+$16.7 & $>10$ & $1.4-3.5$ & -- & \citealt{ahq22} \\
G181.1$+$9.5 & $< 16000$ & $\sim 1.5$ & $0.9-1.5$ & \citealt{krf17} \\
\hline
\end{tabular}

Notes. $^a$ Estimated by dividing the measured post-shock density by the compression ratio (r$_c$=4). $^b$ Estimated by dividing the observed radius and the shock velocity inferred from proper motion for a distance of 1 kpc
\label{tab:high_lat_SNRs}
\end{table*}

Calvera's SNR joins a growing class of SNRs discovered at high Galactic latitude (see Table~\ref{tab:high_lat_SNRs}), each providing a window into SNe evolution in the Galactic halo.
Hoinga (G294.5+24.5) and G354.0–33.5 exhibit low densities ($n_0\sim 0.1-0.4$ cm$^{-3}$) and ages spanning tens to hundreds of kyr, similar to Calvera's SNR middle-aged status but on the more evolved end.
G116.6–26.1 and G70.0–21.5 reside at distances $>$2 kpc and display extremely low densities ($n_0 <10^{-3}$ cm$^{-3}$), highlighting the diversity of halo environments; Calvera's SNR’s higher density underscores localized wind-shaped CSM.
G017.8+16.7 and G181.1+9.5 resemble Calvera's SNR in both evolutionary stage and ambient interaction, suggesting that moderately dense pockets in the halo can substantially influence SNR evolution.
Together, these remnants reveal that, despite their common origin, high-latitude SNRs span a wide range of explosion energies, ambient densities, and evolutionary phases. Calvera's SNR with its prominent O emission line stemming from shocked CSM and $\gamma$-ray counterpart, emphasizes the role of progenitor winds and hadronic processes in shaping the multiwavelength emission of middle-aged SNRs outside the Galactic plane.

\section{Summary and conclusions}
\label{sect:conc}

Despite the significant reduction in the exposure time caused by high instrumental background, our \xmm\ observations provided a significant advance compared to the previous X-ray data that partially covered this sky region. In particular, we could obtain the first measure of the temperature of the thermal plasma and an estimate of the density in the high Galactic latitude Calvera's SNR. Our main findings are the following:

\begin{itemize}
\item The \xmm\ spectra show that the brightest diffuse feature consists of a soft thermal emission ($kT \sim 0.15$ keV) with significant oxygen and marginal nitrogen emission lines, consistent with CIE conditions in a plasma of electron density in the range $n_{\rm{e}} \sim 0.1 - 0.7$ cm$^{-3}$ (depending on the distance). We interpret this emission as resulting from the shocked CSM.
\item Sedov–Taylor and ionization timescale estimates both converge on an age of 10--20 kyr and a distance of $\sim$ $4--5$ kpc for a classical explosion energy of $10^{51}$ erg. A slightly lower energy would make the distance fully compatible with that derived for the Calvera pulsar.
\item Deep H$\alpha$ imaging performed with the TNG/DOLORES confirms the presence of a ``smudge'' filament at the CSM boundary, indicating interaction of the shock with a denser environment.
\item The brightest X-ray region is very close to the maximum in the $\gamma$-ray emission revealed with \fermi/LAT; the density derived from the X-ray analysis, higher than typical halo values, the patchy morphology of the $\gamma$-ray emission and the presence of the H$\alpha$ smudge indicate that the CSM is dense enough to enable efficient pp interactions, suggesting that the hadronic scenario could be responsible for the observed $\gamma$-ray emission, although a scenario in which both leptonic and hadronic channels are at work cannot be excluded, particularly in the outskirts of Calvera's SNR.
\item We also detect a compact but extended Clump with harder emission, equally well fitted by a thermal ($kT \sim 1.7$ keV) or nonthermal ($\Gamma \sim 2.5$) model, whose relation to Calvera's SNR is unclear.

\end{itemize}

Overall, these results are consistent with the origin of the peculiar X-ray pulsar Calvera in the supernova explosion of a massive star that occurred at a large height above the Galactic plane. Our estimated distance of $\sim$4--5 kpc for Calvera's SNR is higher than that derived for the Calvera pulsar from the modeling of its thermal emission ($\sim$ 3.3 kpc, \citealt{mrt21}), but the two values can be reconciled considering the systematic uncertainties inherent in both derivations. Certainly, a distance of the order of $\sim$4--5 kpc is more demanding in terms of the required velocity and lifetime of the putative run-away progenitor of Calvera,  This suggests a slightly less energetic explosion, with $E_{\rm{SN}}$  $\sim$(7--8) $\times 10^{50}$ erg.

The multi-wavelength picture of Calvera's SNR is quite complex and definitely deserves deeper observations and a more complete mapping of the whole region. A good characterization of the properties of Calvera's SNR and other remnants far from the Galactic plane holds great potential to study the physics of shocks evolving in tenuous media.

\begin{acknowledgements}
This work made use of observations made with the Italian Telescopio Nazionale Galileo (TNG) operated by the Fundación Galileo Galilei (FGG) of the Istituto Nazionale di Astrofisica (INAF) at the Observatorio del Roque de los Muchachos (La Palma, Canary Islands, Spain). E.G. and V.S. acknowledge support from the INAF Minigrant RSN4 “Investigating magnetic turbulence in young Supernova Remnants through X-ray observations". S.O., M.M., F.B., and V.S. acknowledge financial contribution from the PRIN 2022 (20224MNC5A) - “Life, death and after-death of massive stars” funded by European Union – Next Generation EU. S.O., M.M., F.B. and E.G. acknowledge financial contribution from the INAF Theory Grant “Supernova remnants as probes for the structure and mass-loss history of the progenitor systems”. M.R. acknowledges the NRRP - funded by the European Union - NextGenerationEU (CUP C53C22000430006). We acknowledge financial support through the INAF grants “Magnetars” and “Toward Neutron Stars Unification”.
\end{acknowledgements}

\bibliography{references}
\bibliographystyle{aasjournal}

\appendix

\section{Cross-match with multi-wavelength catalogues}
\label{app:pointsources}
The output of \texttt{edetect\_chain} is a list of 19 detected sources, including the Calvera pulsar (see Fig.~\ref{fig:pointsources}).
Five sources are spatially coincident with the diffuse emission reported by \citet{zhi11}. We performed a cross-match of these X-ray sources with several catalogs, such as 2MASS, Gaia DR3, SDSS and PanSTARRS, by using a 2$''$-radius circle. Only one match, SRC 13, was found with a PanSTARRS object of magnitude 21.5, likely an active galaxy.
We also compared the list of detected sources with NED, with a radius of 3$''$, finding counterparts for seven sources: 1, 3, 4, 7, 9, 11, 16, marked in yellow in Fig.~\ref{fig:pointsources}.
All the matches found, except for the Calvera Pulsar, also show a WISEA counterpart. We also noticed an offset of roughly 2$''$ between the coordinates found for the Calvera pulsar by \texttt{edetect\_chain} and its position estimate from precise proper motion measurement reported in \citet{rmh24}. Moreover, a comparison between the coordinates inferred from the source detection algorithm and the known sources reveals a similar offset.
In any case, given the spatial resolution of the images this small astrometric mismatch has no significant effect on our analysis of the diffuse emission.

We performed also a source detection on the \chandra/HRC pointings of this sky region (obsID 15806, 26623, 29305, see \citealt{rmh24} for details).
We made use of the \texttt{fluximage} and \texttt{wavdetect} tasks of the \chandra\ Interactive Analysis of Observations software (CIAO, version 4.16), setting default energy range, a binning factor of 4, and an encircled energy fraction of 0.75 at 1 keV.
The sources detected above a threshold of $5\sigma$ in each observation are shown with green crosses in Fig.~\ref{fig:pointsources}.
Sources 1, 2, 4, 5, 7, 14 show a positive cross-match between \xmm\ and \chandra, confirming their point-like origin. The region where \xmm\ detected diffuse X-ray emission is at the edge of the \chandra\ FoV, reducing our detection sensitivity. Nonetheless, we found no sources down to
a significance of $2.7\sigma$; the only detected source (green circle in Fig.~\ref{fig:pointsources}) does not match any of the \xmm\ sources.
We therefore conclude that sources 8, 11, 12, 13, 17, 18 are bright knots of a diffuse component rather than point-like sources.

\begin{figure}[!htb]
 \centering
 \includegraphics[width=0.9\linewidth]{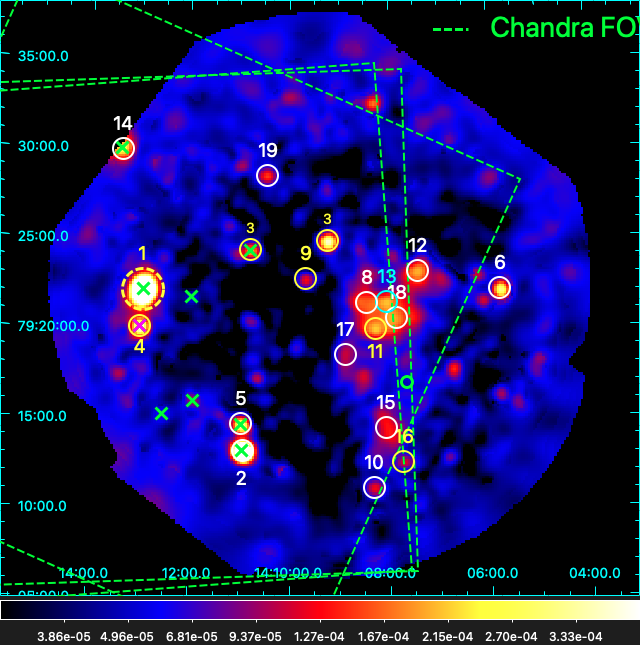}
 \caption{Background-subtracted, vignetting corrected and mosaicked count-rate image in the 0.3--2.5 keV energy range. Circles mark the sources identified by the \texttt{edetect\_chain} tool. Dashed circle, SRC 1, is the Calvera pulsar; yellow circles identify sources having a NED counterpart; cyan circle mark the sources with a PanSTARRS counterpart; white circles are regions with no known counterpart in the considered catalogs. The green dashed lines and crosses mark the \chandra\ FoV and detected sources, respectively.}
 \label{fig:pointsources}
\end{figure}

\end{document}